\documentclass[conference]{IEEEtran}
\IEEEoverridecommandlockouts
\usepackage{cite}
\usepackage{amsmath,amssymb,amsfonts}
\usepackage{algorithmic}
\usepackage{graphicx}
\usepackage{textcomp}
\usepackage{xcolor}
\usepackage{algorithm}
\usepackage{enumitem}
\usepackage{dsfont}
\newcommand{\sys}{\textsf{RepChain}}
\graphicspath{{figures/}}
\def\BibTeX{{\rm B\kern-.05em{\sc i\kern-.025em b}\kern-.08em
    T\kern-.1667em\lower.7ex\hbox{E}\kern-.125emX}}
\begin{document}
\title{\sys: A Reputation-based Secure, Fast and High Incentive Blockchain System via Sharding}

\author{ \IEEEauthorblockN{Chenyu Huang\IEEEauthorrefmark{1}\IEEEauthorrefmark{2}, Zeyu Wang\IEEEauthorrefmark{1}\IEEEauthorrefmark{2}, Huangxun Chen\IEEEauthorrefmark{1}, Qiwei Hu\IEEEauthorrefmark{3}, Qian Zhang\IEEEauthorrefmark{1}, Wei Wang\IEEEauthorrefmark{3}, Xia Guan\IEEEauthorrefmark{4}}
\IEEEauthorblockA{
\IEEEauthorrefmark{1}\textit{Computer Science and Engineering, Hong Kong University of Science and Technology, Hong Kong}\\
\IEEEauthorrefmark{3}\textit{School of Electronic Information and Communications, Huazhong University of Science and Technology, Wuhan, China}\\
\IEEEauthorrefmark{4}\textit{Oasis Futrue (Shenzhen) Holdings, Shenzhen, China}\\
\textit{E-mail: chuangak@connect.ust.hk, {zwangas, hchenay}@cse.ust.hk, hu2013qiwei@hust.edu.cn}\\
\textit{ qianzh@cse.ust.hk, weiwangw@hust.edu.cn, gx@oasisfuture.com}\\
\IEEEauthorrefmark{2}\textit{Co-primary Authors}\\
}
}
\maketitle

\begin{abstract}
In today's blockchain system, designing a secure and high throughput blockchain on par with a centralized payment system is a difficult task. Sharding is one of the most worthwhile emerging technologies for improving the system throughput while maintain high security level. However, previous sharding related designs have two main limitations: Firstly, the throughput of their random-based sharding system is not high enough as they did not leverage the heterogeneity among validators. Secondly, to design an incentive mechanism to promote cooperation could incur a huge overhead on their system. In this paper, we propose \sys, a reputation-based secure and fast blockchain system via sharding, which also provides high incentive to stimulate node cooperation. \sys\ utilizes reputation to explicitly characterize the heterogeneity among the validators and lay the foundation for the incentive mechanism. We propose a new double-chain architecture which includes transaction chain and reputation chain. For transaction chain, a Raft-based synchronous consensus that can achieve high throughput has been presented. For reputation chain, the synchronous Byzantine fault tolerance that combines collective signing has been utilized to achieve a consensus on both reputation score and the related transaction blocks. It supports a high throughput transaction chain with moderate generation speed. Moreover, we propose a reputation-based sharding and leader selection scheme. To analyze the security of \sys, we propose a recursive formula to calculate the epoch security within only $\mathcal{O}(km^2)$ time. Furthermore, we implement and evaluate \sys\ on the Amazon Web Service platform. The results show our solution can enhance both throughout and security level of the existing sharding-based blockchain system.

\end{abstract}

\begin{IEEEkeywords}
Blockchain, Reputation, Sharding 
\end{IEEEkeywords}


\section{Introduction}

The Internet of Things (IoT) is a promising technique to enhance the connection between all physical objects in the world, which was 4.9 billion in 2015 and expected to be 25 billion by 2020~\cite{rivera2014gartner}. However, IoTs still face many security issues such as point of failure, DDoS, Sybil attack and \textit{etc}~\cite{lu2018internet}. In recent years, the development of blockchain technology has shown great potential for its integration with IoTs to address the aforementioned security issues~\cite{novo2018blockchain, yang2018blockchain, huang2019b}.  Technically, blockchain is a decentralized and public digital ledger which records data in a large number of distributed nodes. In ideal cases of such an intermediary-free system, consensus, \textit{i.e.} an agreement on data is expected to be achieved efficiently by the honest majority (\textit{i.e.}, high throughput) and will be resistant to retroactive modification by malicious users (\textit{i.e.}, high security). Existing blockchain systems are still far from meeting the above expectations. For example, Bitcoin and Ethereum can only handle 7 and 15 transactions per second respectively~\cite{ethtps}, which barely meet the requirement of IoTs. 



The fundamental limitation is attributed to a global consensus requirement, \textit{i.e.}, all data should be validated by every validator. Thus, the validation workload increases with the growing number of validators, but the system capacity remains unchanged. Sharding is an intuitive solution to tackle this limitation, where the validators are separated into several groups so that transactions can be processed in parallel to boost throughput. 
In the state-of-the-art literature, several sharding-based protocols (RSCoin~\cite{danezis2015centrally}, Elastico~\cite{luu2016secure}, OmniLedger~\cite{kokoris2017omniledger}, RapidChain~\cite{zamani2018rapidchain} and Monoxoide\cite{wang2019monoxide}) have been proposed to address the trade-off between throughput and security.

However, we argue that a practical blockchain system should take validator heterogeneity and incentive mechanisms into design consideration to fully exploit the potential of sharding techniques. 
Firstly, most existing works regard validators to be the same except for the distinction of honest/malicious attributes. 
However, validators in a practical system present such a difference in terms of computing capability, communication bandwidth and historic behaviors. Thus, under random sharding~\cite{danezis2015centrally,luu2016secure,kokoris2017omniledger,wang2019monoxide} or simple balanced random sharding~\cite{zamani2018rapidchain} protocols, those less-competent validators become bottlenecks and hamper system throughput.
Secondly, an incentive mechanism is crucial for validator activation and retention. In a practical blockchain system, it is undesirable to rely on a trusted centralized bank~\cite{danezis2015centrally} to allocate rewards. Simply allocating rewards to a shard leader or block miner (\cite{luu2016secure, wang2019monoxide}) would result in income variance, \textit{i.e.}, the rewards of validators with modest capabilities have high variance, which may dampen validators' enthusiasm for participation.

In this paper, we introduce `reputation' in the context of the sharding system to jointly address the above two issues in practical systems. 
A reputation established on historical behaviors is the cornerstone of many trust systems. For example, merchants in the markets build up their reputations on long-term fair trading to earn trust from customers~\cite{selnes1993examination, resnick2002trust} and; nodes in Peer-to-Peer systems establish their reputations on active participation in file sharing to obtain other nodes' cooperation~\cite{xiong2004peertrust, zhou2007powertrust}. In recent years, such a concept has infiltrated the blockchain~\cite{yu2019repucoin, huang2019b, bahri2019trust}. However, their systems are not designed for sharding protocols. Thus, to fill this gap, we propose \sys\ to integrate reputation with a sharding-based blockchain to serve the following purposes:

\begin{itemize}
	\item Reputation boosts the system throughput by helping elect a highly capability leader in the shard leader selection. 
	\item Reputation enhances the system's security by helping to balance multiple shards to have similar proportions of active, inactive, honest and malicious validators so that it is more difficult for attacker to take control of one shard. 
	The basic idea to achieve this goal is enforcing different shards to have the same sized and similar total reputation scores during the sharding procedure. 
	\item Reputation greatly provides great incentive for the validators to do their best by helping establish a reputation-based reward scheme. Therefore, every devoted contribution will be recognized and an honest and competent validator who contributes more will be allocated more rewards. 
\end{itemize}



To realize the desired system, two challenges should be addressed. 
Firstly, we should design an efficient consensus protocol for transactions, which explicitly considers heterogeneity among validators to avoid those with lower processing capabilities and more security breaches becoming the bottlenecks.   
Secondly, we should design an efficient consensus protocol for reputation scores. On the one hand, such a consensus should be reached to gain all the benefits from reputation on throughput, security level and incentive mechanism. On the other hand, we should keep the system overhead brought by reputation score at an acceptable level without impairing its benefits. 

Thus, \sys\ novelly proposes the double-chain architecture, \textit{i.e.}, maintains transaction chain and reputation chain separately, to address the  above challenges. For transaction chain, we introduce a synchronous Raft consensus for efficient transaction blocks generation.
Compared with other works, our scheme eases the workload of the least capable validator and incentivizes the competent shard leader to contribute more so as to improve the system's throughput. An honest and competent validator has more chance of being selected as a shard leader, and the reputation-based scheme will ensure that being honest is the best strategy. 
In terms of reputation chain, a synchronous Byzantine fault tolerance consensus~\cite{zamani2018rapidchain} has been used to prevent a Byzantine attack. 
To balance the benefits and overheads of such a consensus, we pack the hashes of multiple confirmed transaction blocks and the reputation scores of all validators calculated based on their behaviors' into one reputation block. 
In this way, a reputation block provides proof of authenticity for multiple transaction blocks. 
Thus, \sys\ can achieve efficient consensus on both reputation scores and transaction chains without incurring too much overhead, \textit{i.e.}, a reputation chain with moderate generation speed can support a high throughput transaction chain.  
It is worth mentioning that existing sharding systems can also benefit from this idea with a little modification on their consensus and sharding scheme to gain the benefits brought about by reputation.

To evaluate \sys, we first provide analysis on the security, throughput and incentive mechanisms. Compared with previous works~\cite{kokoris2017omniledger, zamani2018rapidchain} that only estimate the failure probability of one epoch, we propose a new recursive formula to give the exact solution in time complexity of $\mathcal{O}(km^2)$. Besides, we implement \sys\ and evaluate it using the Amazon Web Service with 900 instances (450 from US West and 450 from US East). This setting is more realistic compared to Elastico that puts all the instances in the US West or Omniledger/RapidChain that put all the nodes in one local area where the latencies are well-controlled.
We use 900 instances to simulate 1800 nodes and set the shard size to 225 to maintain a high security level.
The results show that our system achieves the throughput of 6852 transactions per second (tps) and the user-perceived latency of 58.2 seconds on average.
We also evaluate our system under three different threat models and various validator capabilities. The results present that our system can enhance the throughput and security level of a sharding-based blockchain system. 

In summary, this paper makes the following contributions:
\begin{enumerate}[noitemsep]
	\item \sys\ is the first to integrate reputation scores with sharding-based blockchain, which explicitly characterizes the heterogeneity among the validators and lays the foundations for the incentive mechanism.   
    \item \sys\ is the first sharding-based system with double-chain architecture, which enables a reputation chain generated via the Byzantine fault tolerance consensus to support a high throughput transaction chain generated via modified Raft.  
	\item We propose the reputation-based sharding and leader selection scheme to boost the system throughput and enhance the security level.
    \item We analyze the security and performance of the system, implement the proposed system and conduct extensive evaluations on Amazon Web Services to validate the effectiveness of our design. 
\end{enumerate}


\section{Related Work}
\label{Backgournd}
In this section, we introduce existing efforts to address the blockchain scalability problem, and elaborate sharding-based blockchains and the reputation-based systems in the state-of-the-art literatures.
\subsection{Blockchain Scalability}
It is well known that scalability is an unsolved issue in existing blockchain systems especially considering the integration of IoT and blockchain~\cite{iwanicki2018distributed,lu2018internet}. 
Considerable effort has been made to tackle this problem from various perspectives. Bitcoin-NG~\cite{eyal2016bitcoin} boosts system throughput by allowing the shard leaders in an epoch to append more blocks than those in Bitcoin. However, this method results in substantially increasing the computation overhead of each node. 
ByzCoin~\cite{kogias2016enhancing} leverages a collective signing technique to reduce the communication complexity of PBFT, but it does not make fundamental changes to PBFT which solve the scalability issue.  
Lightning network~\cite{poon2016bitcoin} and Bolt~\cite{green2017bolt} both resort to off-chain payment, which allows participants to transfer micro-payments through a payment channel without appending it onto the main chain. However, off-chains generally do not have the same security guarantees as the main chain, which results in vulnerability to various attacks~\cite{lnddos,ruffing2015liar}. 

\subsection{Sharding-based Blockchain}
A sharding-based blockchain basically distributes transactions to different shards. It is an intuitive solution to improve scalability, as the total throughput is equal to the product of in-shard throughput and the number of shards. This idea has attracted the attention of researchers and a few high throughput and secure systems have been built upon it. 
RSCoin~\cite{danezis2015centrally} leverages sharding to build a centrally banked system, where a simple Two Phase Commit (2PC) has been utilized between the user and a set of mintettes from one shard. 
Elastico~\cite{luu2016secure} proposed the first sharding-based permissionless blockchain with BFT consensus. It achieves a near-linear computational scalability and tolerates corruptions of 1/4 nodes. 
OmniLedger~\cite{kokoris2017omniledger} also adopts sharding, which selects a leader via a verifiable random function (VRF) and utilizes a variant of ByzCoin~\cite{kogias2016enhancing} to improve throughput.
RapidChain~\cite{zamani2018rapidchain} considers sharding in the context of synchronous protocols, which improves throughput and is resilient to the corruption of 1/3 nodes. 
Monoxide~\cite{wang2019monoxide} proposes asynchronous consensus zones that shard the Proof-of-Work (PoW) blockchain. They propose eventual atomicity to ensure transaction atomicity across zones, and Chu-ko-nu mining to ensure effective mining power within one zone.
Dang\textit{et al.}\cite{dang2019towards} propose a sharding blockchain based on trusted hardware, Intel SGX, to achieve high performances for both consensus and shard formation protocols. 

\subsection{Reputaion and Blockchain}
In a traditional P2P network, it is common to leverage reputation as an incentive mechanism~\cite{xiong2004peertrust, zhou2007powertrust}. 
In recent years, the reputation concept has penetrated blockchain. 
CertChain~\cite{chencertchain} proposed a Dependability-rank based consensus and incentive mechanism, which takes economic benefits and misbehaviors into consideration. 
However, their design is tailored to certifying authorities (CA) but not to address the scalability issue. 
B-IoT~\cite{huang2019b} and PoT~\cite{bahri2019trust} both propose credit-based PoW systems to reduce the high computing cost of PoW. B-IoT~\cite{huang2019b} built on Directed Acyclic Graph(DAG), where valid transactions are possible to be falsely appended to parasite chains~\cite{wang2019sok}. PoT~\cite{bahri2019trust} generates the reputation based on the trust network that is reported by every node. However, to defend themselves from attacks, they assume a trust seeding where a set of trusted nodes are identified and the trust network evolves around it.
RepuCoin~\cite{yu2019repucoin} tried to integrate reputation with blockchains. They  proposed a reputation-based weighting scheme consensus to avoid high computation cost as in PoW. 
However, their reputation scores are based on the total amount of valid works from the very beginning of the chain without decay. It could result in a severe monopoly problem and double spending attacks if high reputation users collude with each other. 
Moreover, it is extremely difficult for new validators to join the consensus group and get the rewards. More seriously, their system could be defeated if an attacker joins in the very beginning. 


\begin{figure*}[!thbp]
	\centering
	\begin{minipage}[t]{0.62\textwidth}
		\includegraphics[width=1\textwidth]{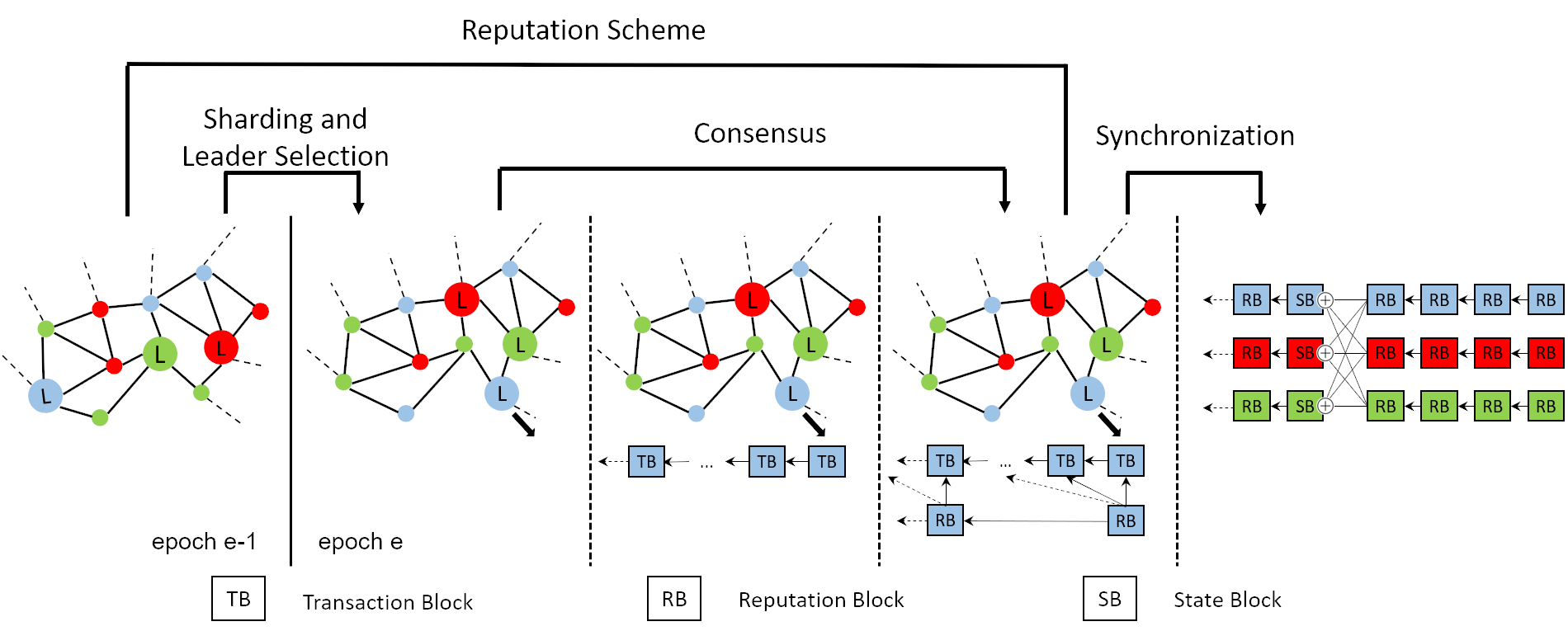}
		\caption{\sys\ Overview}
		\label{fig_overview}	
	\end{minipage}
	\hspace{1.2ex}
	\begin{minipage}[t]{0.35\textwidth}
		\includegraphics[width=1\textwidth]{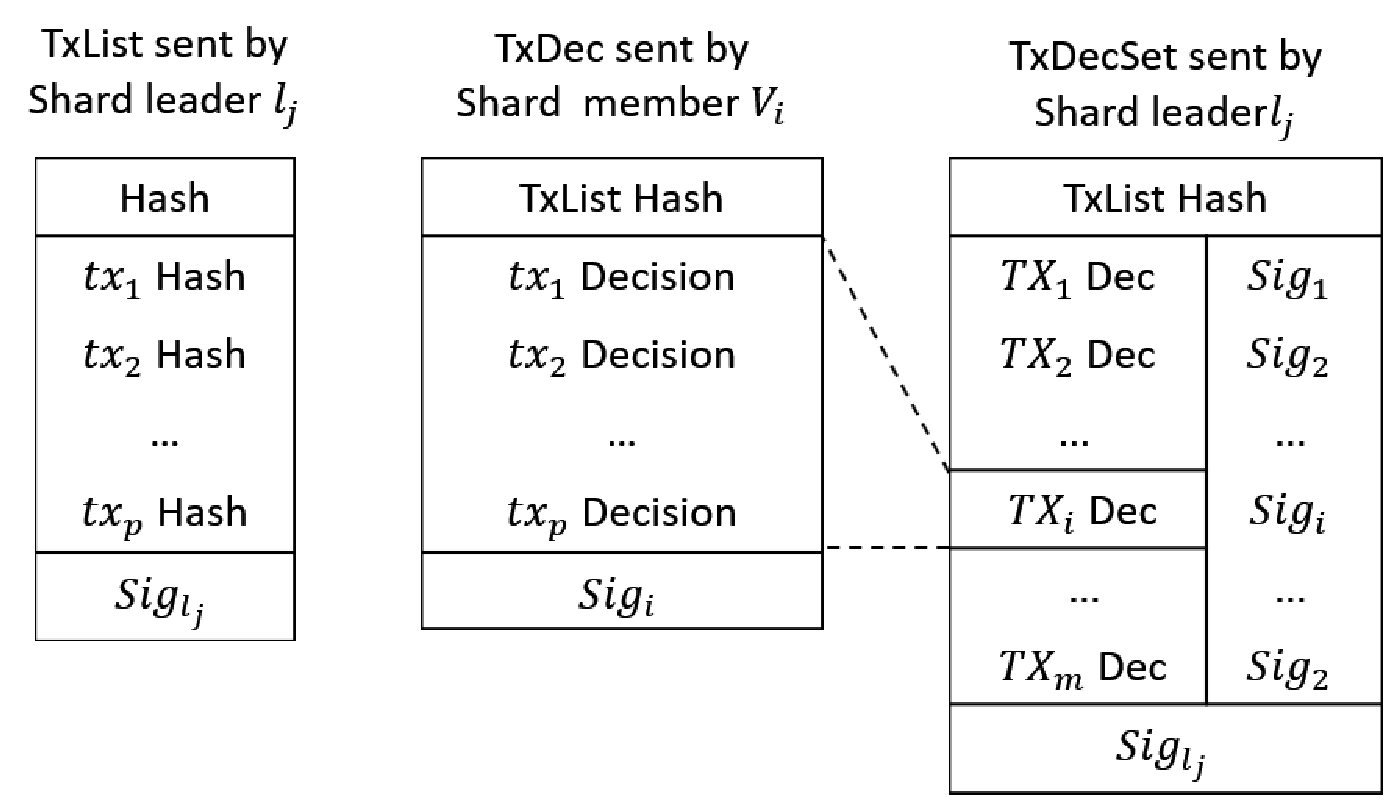}
		\caption{Data structure of TxList, TxDec and TxDecSet }
		\label{fig_tx} 
	\end{minipage}
		\vspace{-4mm}
\end{figure*}

\section{Model and Overview}
\label{Overview}
This section introduces the system model, network model, threat models and the overview of \sys\ respectively.

\subsection{System Model}
An \emph{epoch} in \sys\ denotes the time interval between events of validator assignment to shards as other sharding blockchains~\cite{luu2016secure, kokoris2017omniledger, zamani2018rapidchain}. 
Without loss of generality, we assume there are $n$ validators $V^n = \{v_1, v_2, ... ,v_n\}$ and $k$ shards $C^k = \{C_1, C_2, ..., C_k\}$ in an epoch $e$. Thus, there are $m=n/k$ validators in each shard, including one leader and $m-1$ members. The reputation score of each validator in epoch $e$ is $r^e_i$.
Each node can join the system by generating the sybil-resistent identity leveraging techniques in~\cite{kokoris2017omniledger} and \cite{zamani2018rapidchain}. Specifically, each node should solve a computationally-hard puzzle on its locally-generated identity (i.e., public key) verified by all other (honest) nodes. 
Clients will send transaction $tx$ to validators, who will generate both a transaction chain and a reputation chain via consensus.

\subsection{Network Model}
Our network model is similar to that in~\cite{zamani2018rapidchain}. 
We assume all messages delivered in the network are authenticated with the sender’s private key.
Moreover, we assume that the connections between honest nodes are well established and the transmission can be finished within duration $\Delta$ in the intra-shard consensus, \textit{i.e.}, the communication channel is synchronous within one shard. To address the poor responsiveness of a synchronous consensus, we adopt the method in RapidChain~\cite{zamani2018rapidchain}, where all the members would agree on a new $\Delta$ every week to achieve long-term responsiveness. Except for the intra-shard consensus, \sys\ have been built on partially-synchronous channels with optimistic exponentially increasing time-outs.
\subsection{Threat Model}
We assume a Byzantine adversary who corrupts less than $f=n/3$ nodes. The corrupted nodes can collude with each other and conduct arbitrary behaviours such as sending invalid information or remaining silent. It is assumed that the adversary corrupts the fixed part of the nodes, which is different from the slowly adaptive attackers in~\cite{kokoris2017omniledger,luu2016secure,zamani2018rapidchain}. 
We argue that in a practical system, validators rarely have the same capabilities and secure protection, \textit{i.e.}, some nodes are easily corrupted while others are more robust. 
Besides, our system only kicks out a node when it misbehaves as the shard leader. Thus, for the adversary's benefit, it is more realistic to control the fixed parts of the nodes and pretend to be good nodes until they get higher reputations or are even selected as shard leaders. They can then suddenly launch an attack on the system. 
Thus, this paper considers three attacker strategies as follows:

\emph{Simple Attack}: Malicious nodes do bad things continuously.

\emph{Camouflage Attack}: A malicious node pretending to be a normal node until it becomes a shard leader. Other malicious nodes within the same shard will support the malicious leader.

\emph{Observe-Act Attack}: The attacker observes the reputation score distribution of the normal nodes. Then the attacker controls the malicious nodes to act and have the same reputation score distribution as the normal one. The purpose of this strategy is to increase the probability that most malicious nodes are grouped into one shard. 

Moreover, we also consider several attacks on reputation scores as follows.

\emph{Self-Promoting Attack}: A malicious leader tries to increase the reputation scores of the malicious validators.

\emph{Slandering Attack}: A malicious leader tries to decrease the reputation scores of honest validators.

\subsection{System Overview}

Our system maintains a double-chain architecture, \textit{i.e.}, a transaction chain and a reputation chain. It has four main components as in Fig.~\ref{fig_overview}: sharding and leader selection, consensus, reputation scheme, and synchronization . 

\emph{Sharding and Leader Selection}: All the nodes are assigned into different shards at the beginning of each epoch, then each shard selects its shard leader. 

\emph{Consensus}: The clients send transactions $tx$s to the shards which are responsible for the input UTXOs (unspent transaction output). Then these shards run an intra-shard synchronous consensus to extend the transaction blockchain at high speed and the reputation blockchain at moderate speed. 
For cross-shard transactions, our system adopts an atomic cross-shard protocol.

\emph{Reputation Scheme}: Given behaviours of all validators by the consensus, they can calculate and reach a consensus on the reputation scores. The reputation scores enhance previous two components by helping to elect a high capability leader in the shard leader selection, and balance multiple shards to have similar proportions of active, inactive, honest and malicious validators. Therefore, all three components work together to provide a high throughput, secure and high-incentive blockchain.

\emph{Synchronization}: At the end of each epoch, each shard generates a state block to conclude the transaction and reputation blockchains. 
The nodes synchronize and update the stored reputation scores based on these state blocks from all the shards, and then the system starts a new epoch.


\section{System Design}
\label{Design}
This section presents the detailed design of \sys. 

\subsection{Sharding and Leader Selection}

In sharding and leader selection, our system maintains four properties as follows:
\begin{itemize}
\item\emph{Randomness}: It is hard to predict the results of the sharding and leader selection. 
\item\emph{Balance}: Each shard has a similar total reputation score, i.e., similar proportions of active and inactive, and honest and malicious validators. It is difficult for malicious users to take control of one shard.
\item\emph{Uniformity}: The results can be validated by each validator locally without too much communication overhead.
\item\emph{Incentive}: For the validators, the higher the reputation, the higher the probability of being selected as the shard leader.
\end{itemize}

When a new epoch $e$ begins, all the validators will be sharded into different groups. For each group, a shard leader will be selected based on their reputation as Alg.~\ref{alg:Shard} shows. Decaying is an important feature in reputation scheme, thus we adopted a sliding window $w$ to calculate the cumulative reputation score $R^w$.
Specifically, the validators use the random generator $RNG$ from the random $seed^e$ to generate random numbers. Such a seed could be generated by the secure distributed bias-resistant randomness generation protocol which is adopted by Omniledger\cite{kokoris2017omniledger} and RapidChain\cite{zamani2018rapidchain}. These protocols have been investigated well when combined with a sharding based blockchain. Thus, it is not the focus of our paper.
$R^{sort}$ denotes the validators' reputation scores sorted into descending order (Line 3). 
Next, each validator is randomly assigned to of a minimum size to maintain the balance property (Line 4-7). 
For shard leader selection, we adopt the following strategy to maintain the incentive, randomness and uniformity property simultaneously. First, the validators with reputation scores higher than the median have chances of being selected as the leader. Each score of these validators will divide a number that is randomly generated based on the same seed $Seed^e$, then the validator with the minimum result is selected as the leader. (Line 11-17) 

\renewcommand{\algorithmicrequire}{\textbf{Input:}}
\renewcommand{\algorithmicensure}{\textbf{Output:}}
\begin{algorithm}[thbp]
	\caption{Sharding and Leader Selection Algorithm}           
	\label{alg:Shard}               
	\begin{algorithmic}[1]                
		\REQUIRE ~~\\                        
		A random $Seed^e$ and the cumulative reputation score $R^w = r_1^w+r_2^w+...+r_n^w$ over previous $w$ epochs;\\
		\ENSURE ~~\\                                    
		The $k$ shards $C = \{C_1, C_2, ..., C_k\}$;\\
		The $k$ leaders $L = \{l_1, l_2, ..., l_k\}$.
		
		\STATE Initialize $C_i = \emptyset$, $L_i = \emptyset$ for each $1\leq i \leq k$.
		\STATE Set the seed of random generator $RNG$ as $Seed^e$.
		\STATE $R^{sort}$ = sort($R^w$) 
		
		\FOR {each $r_i^{sort}$ in $R^{sort}$ ($r_i^{Sort}$ is validator $v_g$'s score)}
		\STATE Find a shard $C_t$ that has the minimum cardinality. 
		\IF {multiple $C_t$ satisfied the requirements}
		\STATE Randomly select one based on $RNG$	
		\ENDIF
		\STATE Assign validator $v_g$ to $C_t$, $C_t = C_t \cup \{v_g\}$.
		\ENDFOR
		
		\FOR {each shard $C_i \in C$}
		\STATE $rm$ = median of the subset of $R^w$ that belongs to $C_i$
		\FOR {each validator $v_j \in C_i$}
		\IF {$rm \leq r_j^w$} 
		\STATE Generate a random float $0 \leq y \leq 1$ from $RNG$.
		\STATE $p_{i,j} = y/r_j^w$
		\ELSE  
		\STATE $p_{i,j} = +\infty$
		\ENDIF
		\ENDFOR
		\STATE $l_i = v_j$ where $p_{i,j} = \min(p_{i,1}, p_{i,2}, ... p_{i,m})$
		\ENDFOR
	\end{algorithmic}

\end{algorithm}

\subsection{Consensus}
\label{sec_consensus}


Similar to Bitcoin, each transaction $tx$ will have a unique identity, a list of input UTXOs and a list of output UTXOs. UTXO, short for unspent transaction output, is the unused coin from a previous $tx$ and contains the signature. The shard $i$ will only store the $tx$ with a specific identity, \textit{i.e.}, the prefix of the identity is equal to $i$\cite{zamani2018rapidchain,wang2019monoxide}. The input UTXOs may come from different shards. Thus, our system also need to handle cross-shard transactions. We denote the shard responsible for the input UTXO as the input shard, and the shard for the output UTXO as the output shard.

Inspired by Raft\cite{ongaro2014search}, Omiledger\cite{kokoris2017omniledger} and the protocol proposed by Ren \emph{et al.}\cite{abraham2017efficient} and RapidChain\cite{zamani2018rapidchain}, we propose a synchronous consensus to achieve high throughput and 1/2 resilience within a shard. Our consensus constructs two chains within each shard: a Raft consensus to generate a transaction chain and a Byzantine fault tolerance consensus to generate the reputation chain.


\begin{figure}[h] 
    \centering
    \includegraphics[width=0.5\textwidth]{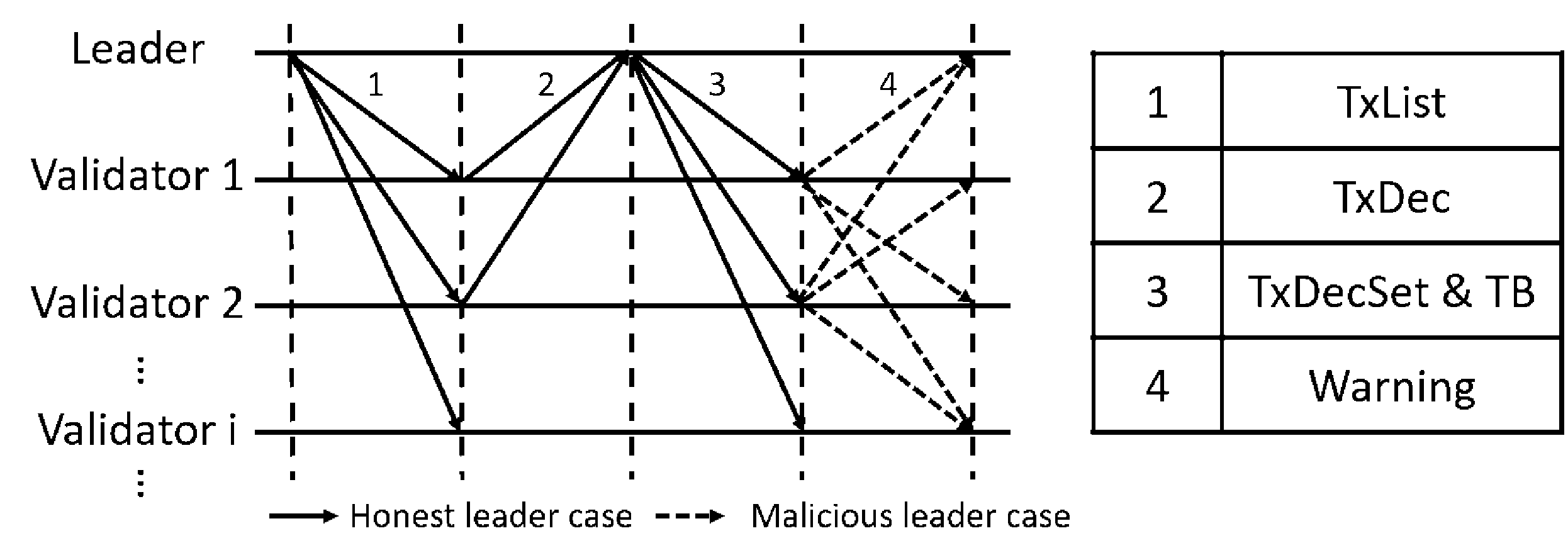}
    \caption{The basic communication pattern of the Raft consensus under honest and malicious leader cases}
    \label{fig_raft} 
\end{figure}

\begin{figure}[h] 
    \centering
    \includegraphics[width=0.5\textwidth]{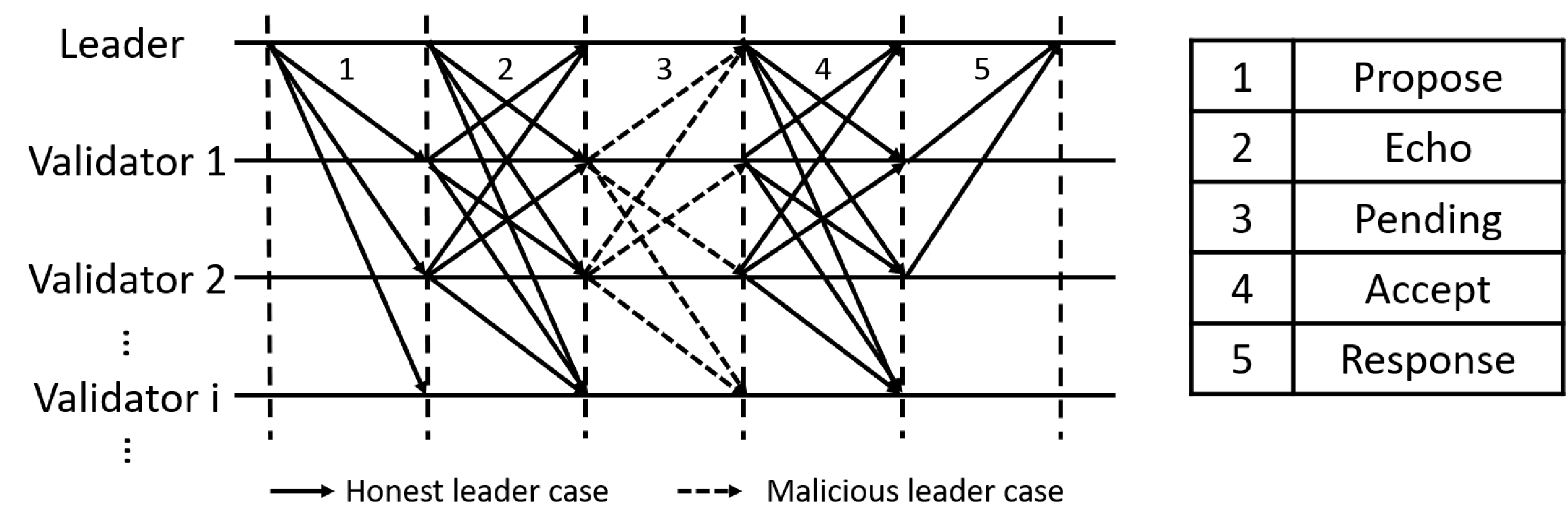}
    \caption{The basic communication pattern of the modified RapidChain's consensus under honest and malicious leader cases}
    \label{fig_raft} 
\end{figure}

\subsubsection{Intra-shard Consensus}
Firstly, the leader generates a transaction chain via Raft consensus. The related data structures are shown in Fig.~\ref{fig_tx}. TxList refers to the transaction list which order the transactions so that every validator can calculate the reputation based on the same batch of transactions. TxDec refers to transaction decisions given by the member, which contains the decisions in the same order as the TxList. TxDecSet refers to the transaction decision set including all the TxDecs from shard members. 
The Fig.~\ref{fig_raft} Illustrates the communication pattern of the Raft consensus. 
(1) The leader will send the TxList to all the validators;
(2) Each validator will validate all the $tx$s in the TxList and make a decision, \emph{Accept}, \emph{Reject} or \emph{Unknown} for every $tx$. \emph{Unknown} is used when the validator cannot handle too many transactions due to the hardware limitations. Then the validators will send the TxDec back to the leader; 
(3) The leader gathers all the TxDecs to the TxDecSet; Meanwhile, the leader can generate the transaction block (TB) based on the TxDecSet. Then the leader will send both the TxDecSet and the TB to all the validators;
(4) If the leader launches a self-promoting or slandering attack by modifying the TxList, TxDecSet or TB (refers to Sec.~\ref{sec_conesensus_warning} for technical detail), the validators will send \emph{warning} to each other and then view-change happens.
After this protocol, the validators can calculate the reputation scores.
The Raft consensus can efficiently generate a transaction block and is more suitable for a high capability leader than the PBFT which need all-to-all communication. However, the Raft consensus cannot prevent a Byzantine attack. Thus, these TBs can only be viewed as candidate TBs and will finally achieve consensus in the second step.

Secondly, \sys\ generates reputation block (RB) via a Byzantine fault tolerance consensus. The structure of reputation block can be viewed in Fig.~\ref{fig_RB} and includes the hashes of transaction blocks to be finally confirmed as well as the reputation of all the validators. 
The consensus from RapidChain\cite{zamani2018rapidchain} can achieve 1/2 resiliency within a shard. However, it is not efficient in a cross-shard transaction since they use the multiple signatures of honest validators as a proof for the transaction validation. Thus, we combine RapidChain consensus with CoSi\cite{kogias2016enhancing} which could aggregate multiple signatures into one signature as Fig.~\ref{fig_raft} shows. Specifically, the consensus of RapidChain contains four synchronous rounds: \emph{propose}, a leader proposes a hash digest $H$ for the consensus; \emph{echo}, validators send the $H$ receives from the leader to all other nodes with tag echo iff the $H$ is correct; \emph{pending}, if an honest validator receives different hashed $H$ and $H'$, it will send the $H'$ with tag pending; \emph{accept},  if an honest validator receives $f+1$ echo of the same and only $H$, it will accept and send $H$ with a tag accept. After the consensus, the honest validator can know whether the leader is corrupt or not.
CoSi also has four rounds: announce, commit, challenge and response. We refer to read the RapidChain\cite{zamani2018rapidchain} and ByzCoin\cite{kogias2016enhancing} for the technical details. The \emph{propose} and \emph{echo} rounds in the consensus can be used for the announce and commit round in the CoSi. Specifically, the leader sends an announcement of the signing of $H$ in the \emph{propose} round. The validators will echo the leader the $H$ and a random secret in the \emph{echo} round. In the \emph{accept} round, the leader send a collective Schnorr challenge and the $H$ with $f+1$ echos. After \emph{accept}, the honest can determine whether the leader is corrupt or not.
Thus, another round of \emph{response} is needed: validators can accept or reject sending the aggregate response to the leader and the leader will generate the collective signature on the RB.


\subsubsection{Warning in Intra-shard Consensus}
\label{sec_conesensus_warning}
In intra-shard consensus, we use a \emph{warning} to detect the malicious behavior of the leader. Specifically, if the leader sends an invalid TxList, TxDecSet or the TB which contains $tx$ less than half of the votes, the $f+1$ honest validator will send a \emph{warning} to the others. Once the $f+1$ \emph{warning} received, view-change happens. 
Malicious leaders can also choose to remove a TxDec of an honest validator from the TxDecSet to decrease its reputation. The victim can send a \emph{warning} with its TxDec to all other validators. 
All the honest validators will add the reputation on the validators. If the malicious leader refuses to add the reputation score of the victim, the hash of RB, $H$, will be different among the malicious leader and the honest validators.
\subsubsection{Cross-shard Protocol}
For cross-shard transactions, we used the Atomix protocol from Omniledger, \textit{i.e.}, the client will help to relay the transactions across the shard with the proof-of-acceptance. In our system, the proof-of-acceptance is the collective signature on RB.

\begin{figure}[h] 
	\centering
	\includegraphics[width=0.5\textwidth]{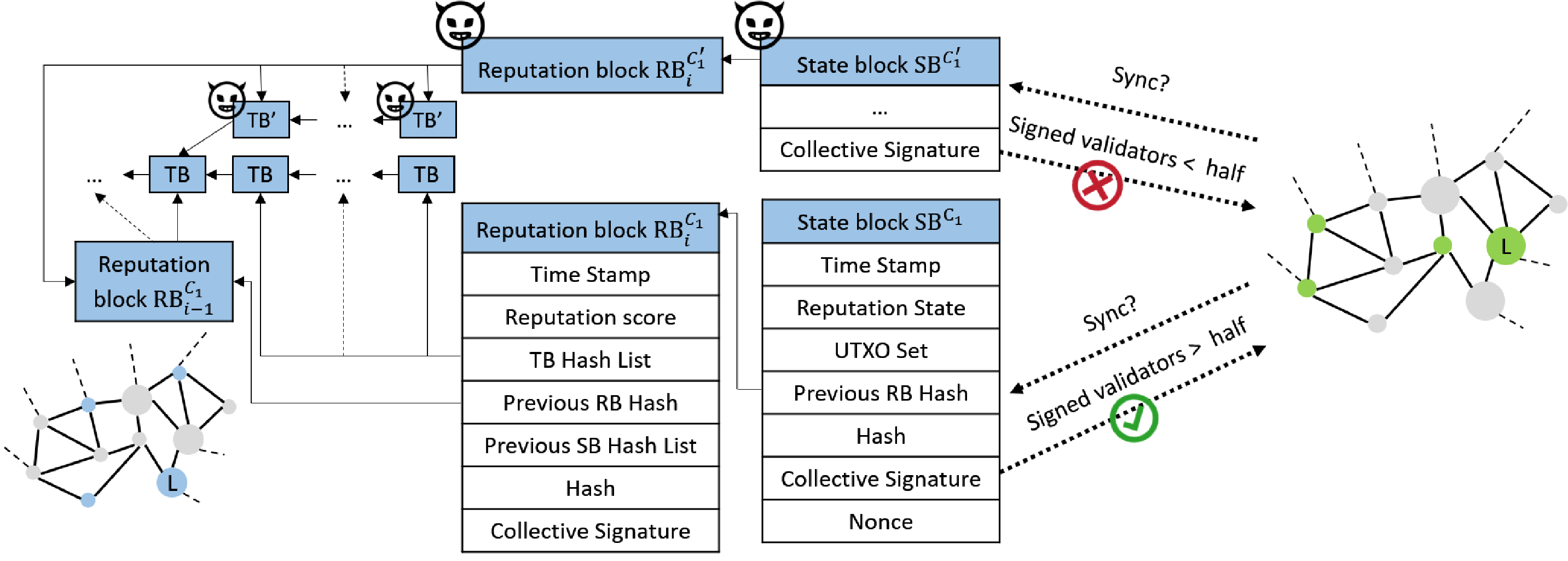}
	\caption{This figure shows the structure of the reputation block (RB) and state block (SB) as well as the working procedure of colective signing in the synchronization process. }
	\label{fig_RB} 
\end{figure}

\subsection{Reputation Scheme}
\label{Sec_ReputationScheme}
Recalling that we use double-chain architecture the previous section explains. In this section, we illustrate the design of the reputation scheme. Our proposed reputation scheme can enhance the security, incentive property and throughput of \sys.  

\subsubsection{Reputation Score Calculation}

At the end of one intra-shard consensus, the reputation scores within the shard can be calculated by all shard members individually and uniformly based on TxDecSet and TB. Inspired by PeerTrust\cite{xiong2004peertrust}, the reputation score $r_i$ of validator $i$ are calculated as follows:
\setlength{\abovedisplayskip}{0pt}
\setlength{\belowdisplayskip}{5pt}
$$r_i = \sum_{j=1}^{l}{S(j)*T(j)}$$
Where $l$ is the number of transactions generated after the previous RB. $T(j)$ is the value of transaction $j$ to prevent the case where some validators are honest on small transactions but dishonest on a large transaction. The scaling factors $S(j)$ are used to reward or penalize different behaviours differentially.  eBay\cite{ebayrepscore} simply gives a reputation score of \{-1, 0, 1\} to correct, unknown and incorrect decisions. However, a validator can still gain a lot of profits even it is dishonest from time to time. 
Thus our system sets different scaling factors for different behaviours to make the punishment for dishonest behaviours larger than the reward for honest ones. Furthermore, an illegitimate $tx$ being passed is more dangerous than a legitimate $tx$ being aborted. Thus, if a validator gives an \emph{"Accept"} while most of the others give a \emph{"Reject"}, the validator will be punished more than when it gives a \emph{"Reject"} while the others give an \emph{"Accept"}. 
For \emph{"Unknown"} decision, the validators neither earn nor lose their reputation scores, since it is believed that these validators generally have lower capacities (CPU, hard disk, and bandwidth etc). 

\subsubsection{Reputation Blockchain}

After the reputation score calculation, the validators utilize collective signing to generate the RB, the structure of which is shown in Fig.~\ref{fig_RB}. RB contains the reputation score the validator owns in this epoch, the confirmed TBs, the previous RB and the collective signature. 
Recalling that in the previous section, the consensus ensures that the honest validators will sign on the RB iff the linked transaction blocks and the reputation scores are correct.
Other shards can check the collective signature, and accept it if more than half of the validators sign on the block. Moreover, the malicious validators can fork the transaction blockchain and the reputation blockchain. However, as shown in Fig.~\ref{fig_RB}, if a malicious validator in shard $C_1$ (blue) forks the blockchain by generating the malicious TBs and RBs, the shard $C_2$ (green) can easily check the collective signature to pick the correct blockchain signed by the honest majority.



\subsection{Synchronization}

In the sharding system, the transactions are separated into different shards. Thus at the end of one epoch, the validators of all shards need to synchronize their reputation blockchain and transaction blockchain to prepare for the next epoch. 
To prevent the huge cost of sending the whole blockchain, the validators launch a round of the RapdiChain's consensus with CoSi to generate a state block (SB). 

The structure of SB is shown in Fig.~\ref{fig_RB}. It contains the overall reputation scores of the validators over the past $w$ epochs and the UTXO set of the clients by this epoch. The UTXO set is the set containing all the UTXO that are generated within this epoch. We also combine the UTXO that belong to one public key. For example, if two UTXOs have the values of $a$ and $b$ respectively and both belong to the public key $PK_x$. The validators will drop the UTXO that has the value of $b$ and adds the value onto the remaining UTXO. The drop principle is to reserve the UTXO with the smaller address. Thus, a validator only needs to download the state block of the other shard instead of downloading all of the data in TB and RB.

\subsection{Double-Chain Architecture}
As mentioned before, we utilize the double-chain architecture to enhance security, throughput and the incentive mechanism. Previous sections detail all of the parts. In this section, we put all the pieces together to show how double-chain architecture works.

The double-chain architecture includes the transaction chain and the reputation chain. \sys\ use Raft consensus to generate a transaction chain at a fast speed. To prevent Byzantine fault, the combination of CoSi and RapidChain's consensus has been used to finally confirm the reputation block and its relevant transaction blocks. The attack on TxList, TxDecSet and TB can be detected early by \emph{Warning}. The Byzantine attack can be finally detected via the modified RapidChain consensus. Thus the TB can only achieve the consensus iff the RB that is linked to it has achieved consensus. Once view-change happens, all the validators will drop the invalid RB and the linked TBs.


\section{System Analysis}
\label{Analysis}
In this section, we analyze the security, the performance and the incentive mechanism of \sys.
\subsection{Epoch Secuirty}
\label{EpochSecurity}
For epoch security, previous work~\cite{zamani2018rapidchain} only estimates the upper bound of the failure probability via hypergeometric distribution since the straightforward calculation for the exact solution requires $\mathcal{O}(m^k)$ time. 
We propose a novel recursive formula to calculate the exact solution in the time complexity of $\mathcal{O}(km^2)$, and prove \sys\ only degrades to the random-based sharding scheme in the worst case.

Firstly, we provide the recursive formula for random-based sharding. 
We denote $F(x,y)$ as the number of safe allocations, where $x$ malicious nodes are assigned to $y$ shards and none of the shards has more than half of the malicious validators. 
An unsafe shard indicates that half of the shard members are malicious nodes. If the above assignment results in any one unsafe shard, we regard such an assignment as a failure, and the failure probability is $P(failure) = 1-F(g,k)/C^g_n$. 
If the $y$-th shard contains $s$ malicious nodes, then the original question is reduced to the number of safe allocations, where $g-s$ malicious nodes are distributed to $y-1$ shards. 
Thus, the equation can be calculated recursively as follows:
\setlength{\abovedisplayskip}{0pt}
\setlength{\belowdisplayskip}{5pt}
\begin{equation}
F(x,y)= \sum_{s=0}^{d}{F(x-s, y-1)C_m^s} 
\label{equ_1}
\end{equation}
where $d = \lfloor \frac{m-1}{2}\rfloor $ and $F(x,1)=C_m^{x}\mathds{1}_{x \leq d} \mathds{n}$. For the case of random-based sharding, given $n=1800$ and $k=8$, $P(failure)=1.25553\mathrm{e}{-07}$. 

Secondly, we prove that the \emph{observe-act attack} degrades the security level as well as the performance of \sys\ to that of the random-based sharding. Since malicious validators maintain the same reputation score distribution as the honest nodes, the possibility of a validator being an attacker or an honest node is independent of the reputation scores. However, according to the calculation above, it is still secure enough.

Thirdly, we prove \sys\ is more secure under \emph{Camouflage Attacker} than random sharding. Given all malicious validators support the malicious leader under \emph{Camouflage Attacker}, they expose themselves due to the view-change and their reputation scores will decrease. Their cumulative reputation score will be lower than honest validators the following $w$ epochs. Assuming $a$ ($a=pk+q,0 \leq q \leq k-1$) malicious nodes are exposed, they will be allocated into $k$ shard equally accroding to our sharding scheme. The allocation is equivalent to distributing $g-pk$ malicious nodes to $k$ shards where each shard already contains $p$ nodes. For clear elaboration, we first consider the remaining $q$ exposed validators will be assigned in the $(k-q+1)$-th to the $k$-th shard in order.
We denote $F(x,y,t)$ as the number of the safe allocations, where $x$ is the number of malicious validators except the exposed ones, $y$ denotes the number of shard and $t$ is the number of exposed malicious nodes. It can be calculated similar to Equation~\ref{equ_1} as follows:
\setlength{\abovedisplayskip}{0pt}
\setlength{\belowdisplayskip}{5pt}
\begin{equation}
F(x,y,t)=\sum_{s=0}^{d -u_l}{F(x-s, y-1, max(t-1,0))C_{m-p-u_l}^s}
\label{equ_2} 
\end{equation}
where $l=k-y$, $u_l=\mathds{1}_{l \in [0, q-1]}$, $v=max \{0, q-l-1\}$ and $F(x,1,0)=C_m^{x}\mathds{1}_{x \leq d} \mathds{n}$. The above recursive equation can be calculated within $\mathcal{O}(km^2)$ time. Since we only consider the situation where the $q$ validators are assigned to the last $q$ shards, the total number of safe allocations should be $C^q_kF(g-a,k,q)$ and the failure probability is $(C^q_kF(g-a,k,q))/(C^q_kC^{g-a}_{n-a})$ when considering all situations. 
Fig~\ref{fig:failure_prob} shows the failure probability with a different $a$. The failure probability decreases with $a$ increasing, \textit{i.e.}, the system is more secure when more malicious nodes are exposed. Thus, the security level of \sys\ under such an attack will be higher than the random sharding scheme. For \emph{Simple Attack}, the attacker can never succeeded since the attackers are always exposed to the system. 


\subsection{Security of Intra-shard Consensus}
\label{TBsecurity}

In Sec.~\ref{EpochSecurity}, we prove that a shard having more than half of the malicious nodes is almost impossible. Based on this, we prove the safety and liveness of our intra-shard consensus.

\subsubsection{Safety of Intra-shard Consensus}
Raft consensus cannot guarantee the safety of the TB owing to it not being a Byzantine fault tolerance consensus. Safety of the RB and the related TBs is achieved by the RapidChain's consensus. Thus, we only need to prove the adding \emph{response} round will not violate the safety property of the original RapidChain's Consensus as follows: As all the honest validators have know whether or not to accept these blocks, the honest validators can then determine whether or not to response to the leader before the \emph{response} round. Thus, the malicious block proposed by the malicious leader cannot gain the response from honest validators to aggregate a collective signature signed by $f+1$ validators.


\subsubsection{Liveness of Intra-shard Consensus}
We use the definition of liveness as the finality for one block which is the same as other works~\cite{zamani2018rapidchain}.  
Recalling that the expectation round for an honest leader is around two in the worst case.
First, we prove the liveness of the Raft consensus on the transaction block. The honest leader will send the valid TxList and since half of the shard members are honest, the TxDecSet can be given. Thus all the honest shard members will accept a valid TB. If the current leader is a malicious one who remains silent or sends malicious information, the predefined timeout and the Byzantine fault tolerance consensus will launch a view-change and a new leader will eventually be selected. Thus a valid TB will finally be proposed by an honest leader. 
Secondly, the liveness of Byzantine fault tolerance is guaranteed by the RapidChain's consensus. The adding of the \emph{response} round will not violate the originally liveness consensus as follows: At the end of the \emph{accept} round, all the honest validators know whether to accept the RB. If the leader is malicious, the view-change happens otherwise the \emph{response} will be finished by all the honest validators.

\subsubsection{Attack on Reputation}
In threat model, we define two attacks on reputation: self-promoting and slandering attacks. In a self-promoting attack, a malicious leader may try to modify the reputation in RB to increase the reputation of malicious validators. However, the reputation score is determined by the TxList and TxDecSet. The inconsistency between them will cause a view-change since the hash $H$ of RB will be different between honest validators and malicious leaders. In a slandering attack, a malicious leader will try to decrease the reputation of the honest validators. As mentioned above, the modification of RB is impossible. Thus the only choice is to remove the TxDec from the honest validators in TxDecSet. However, the \emph{warning} with its TxDec ensures the honest validators add the reputation of the victim  as Sec.~\ref{sec_consensus} shows.

\begin{figure*}

    \begin{minipage}[t]{0.235\textwidth}
    \centering
    \includegraphics[width=1\textwidth]{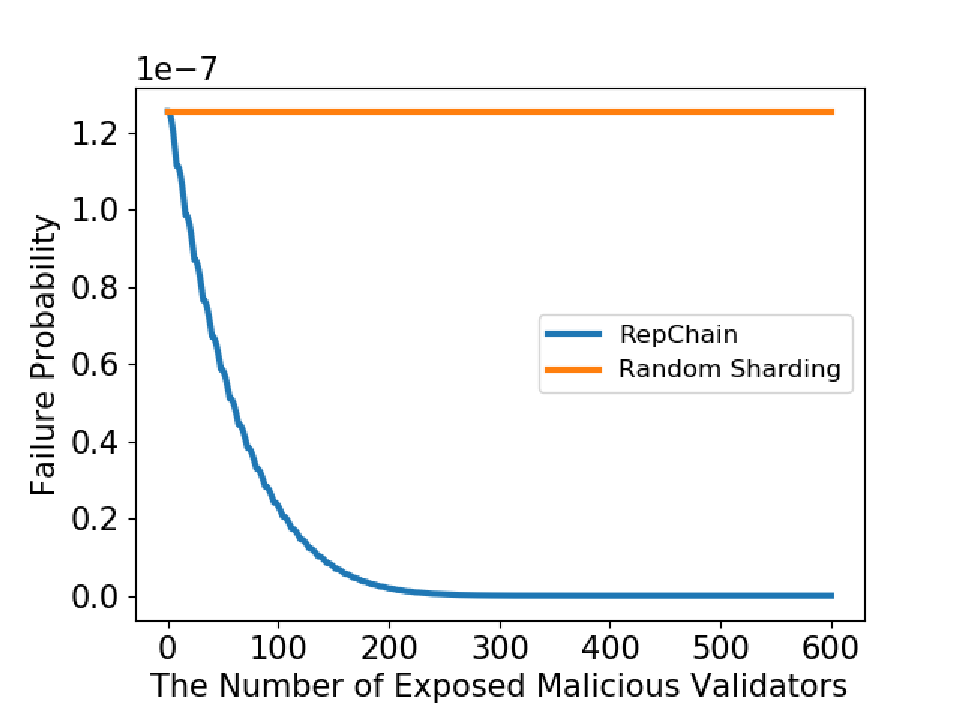}
    \caption{The failure probability of \sys\ with a different number of exposed malicious validators}
    \label{fig:failure_prob}
    \end{minipage}
    \hspace{0.15ex}
    \begin{minipage}[t]{0.235\textwidth}
    \centering
    \includegraphics[width=1\textwidth]{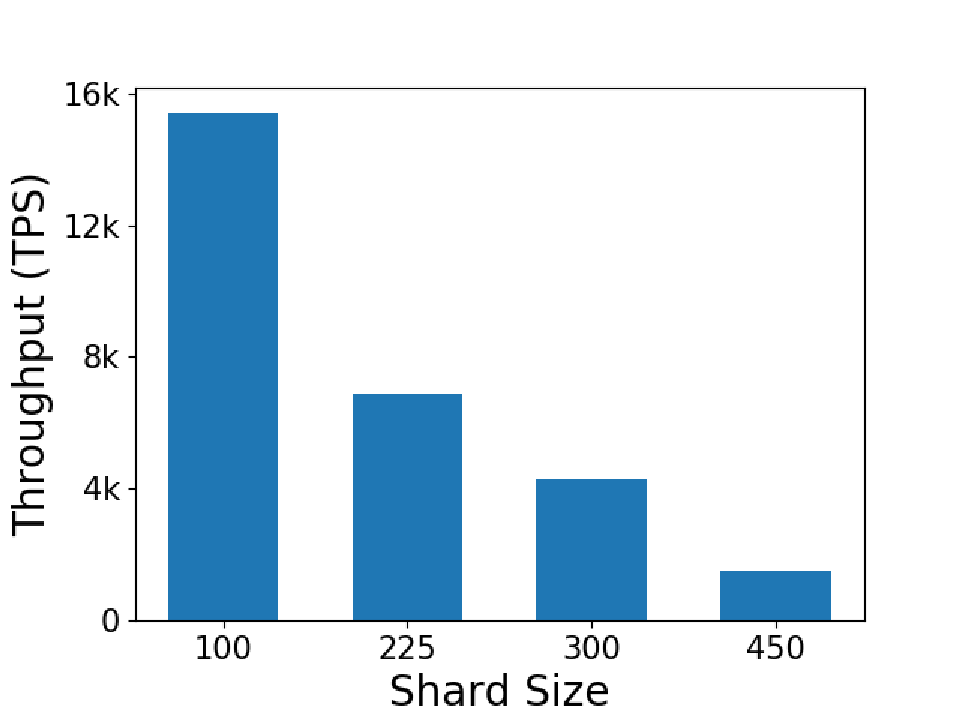}
    \caption{The average throughput of the transactions for different sized shards.}
    \label{fig:TPCshardsize}
    \end{minipage}
    \hspace{0.15ex}
    \begin{minipage}[t]{0.235\textwidth}
    \centering
    \includegraphics[width=1\textwidth]{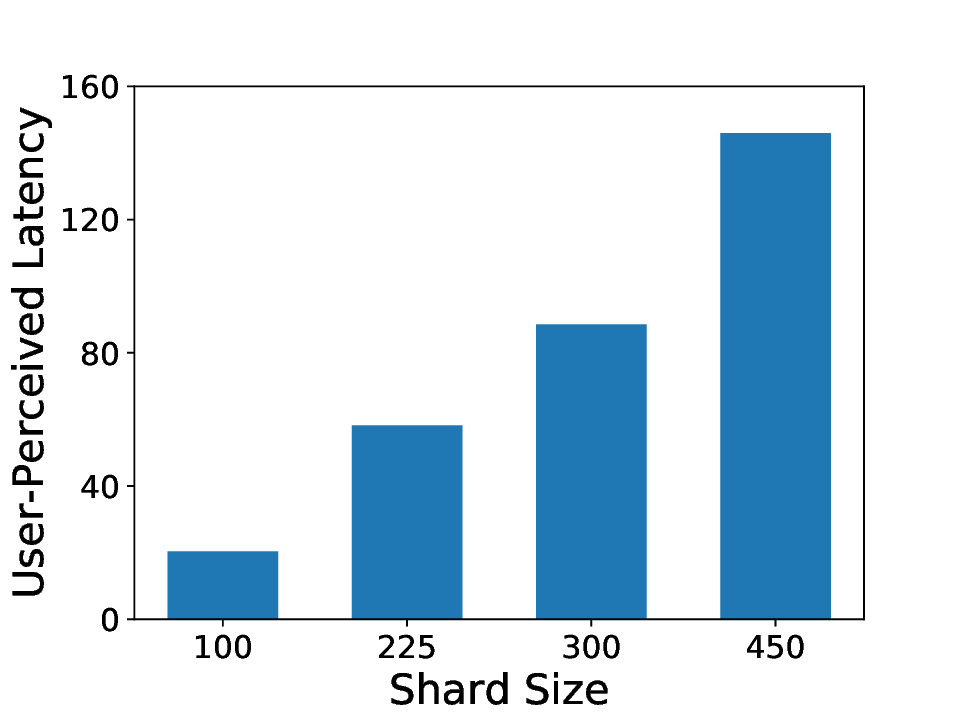}
    \caption{The average user-perceived latency of a transaction for different sized shards}
    \label{fig:delayshardisze} 
    \end{minipage}
    \hspace{0.15ex}
    \begin{minipage}[t]{0.235\textwidth}
    \centering
    \includegraphics[width=1\textwidth]{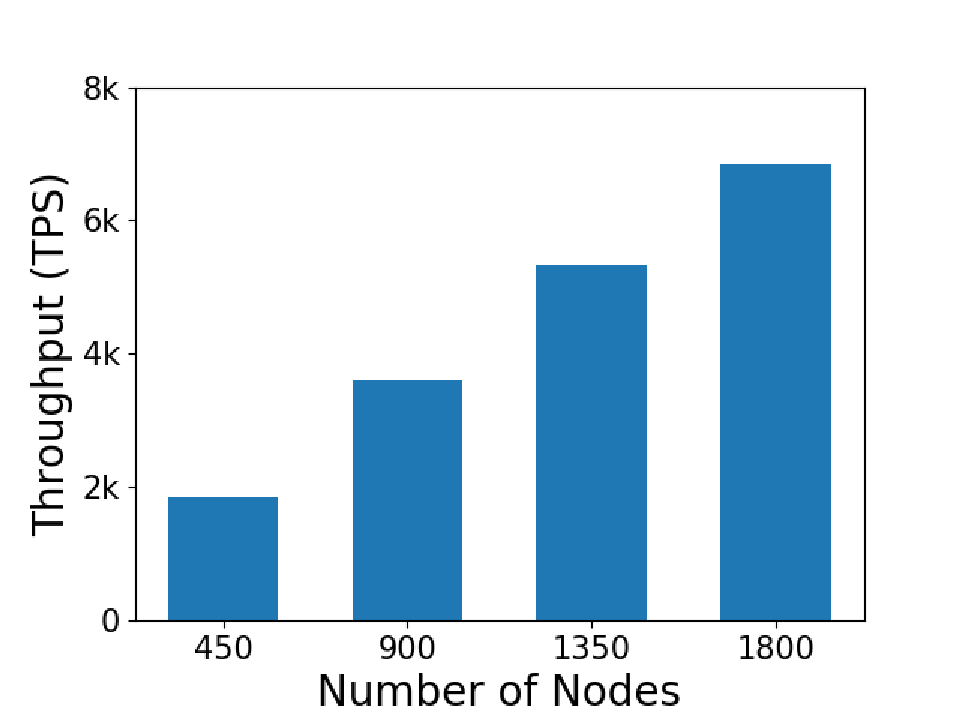}
    \caption{Throughput scalability}
    \label{fig:TPCscalability} 
	\end{minipage}
   \vspace{-4mm}   
\end{figure*}    

\begin{figure*}
	\begin{minipage}[t]{0.235\textwidth}
		\centering
		\includegraphics[width=1\textwidth]{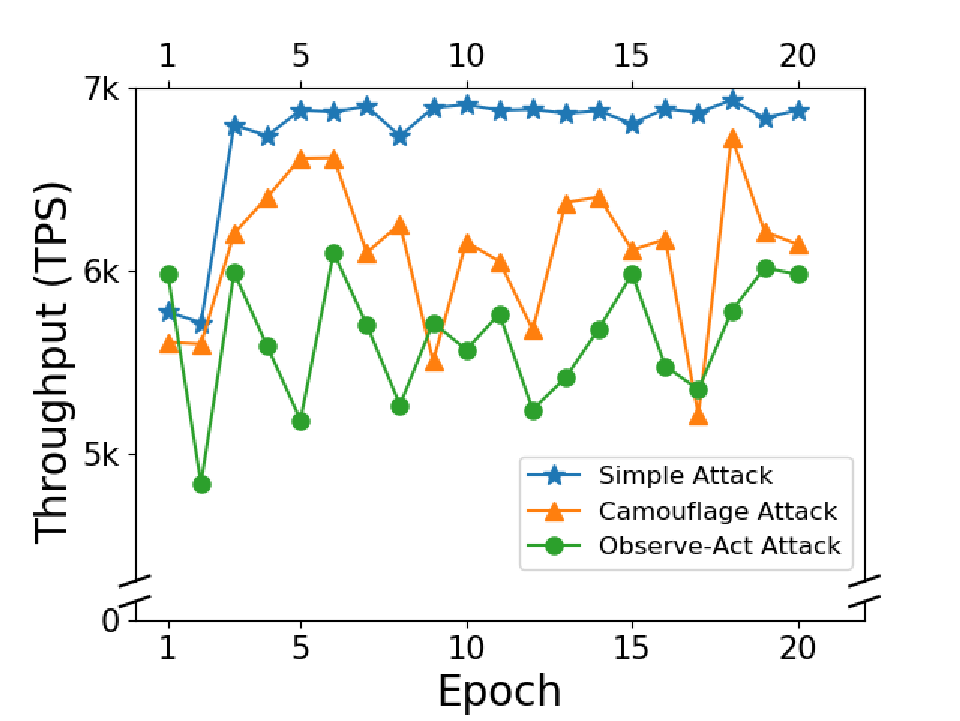}
		\caption{The average throughput for three attack models}
		\label{fig:security} 
	\end{minipage}%
	\hspace{0.15ex}
	\begin{minipage}[t]{0.235\textwidth}
		\centering
		\includegraphics[width=1\textwidth]{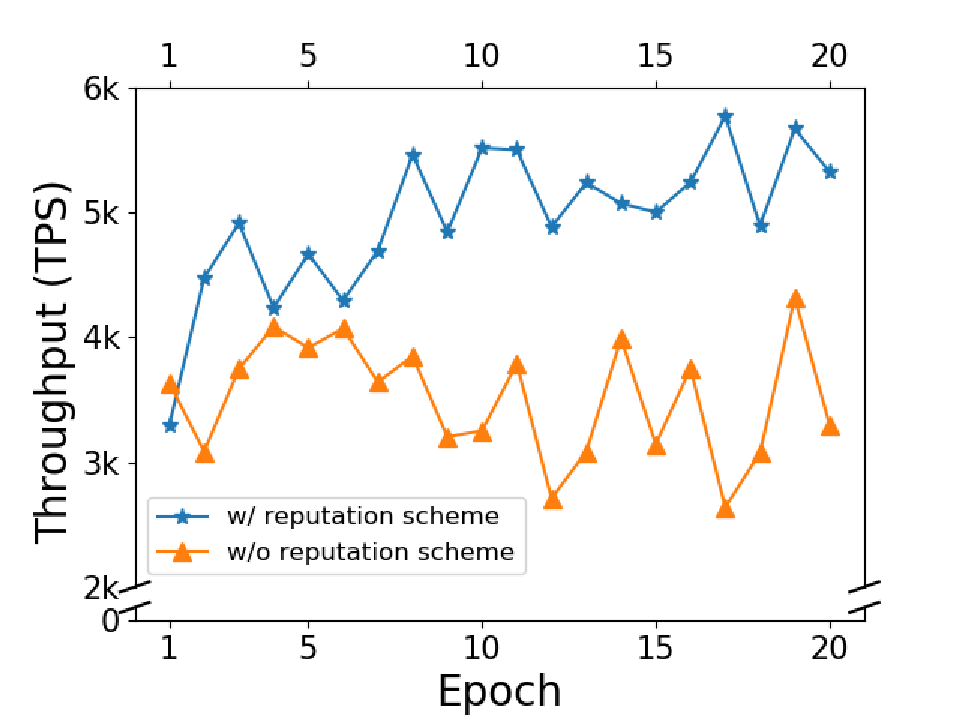}
		\caption{The average TPS under different validators' abilities w/ and w/o reputation scheme.}
		\label{fig:BandwidthTPS}
	\end{minipage}
	\hspace{0.15ex}
	\begin{minipage}[t]{0.235\textwidth}
		\centering
		\includegraphics[width=1\textwidth]{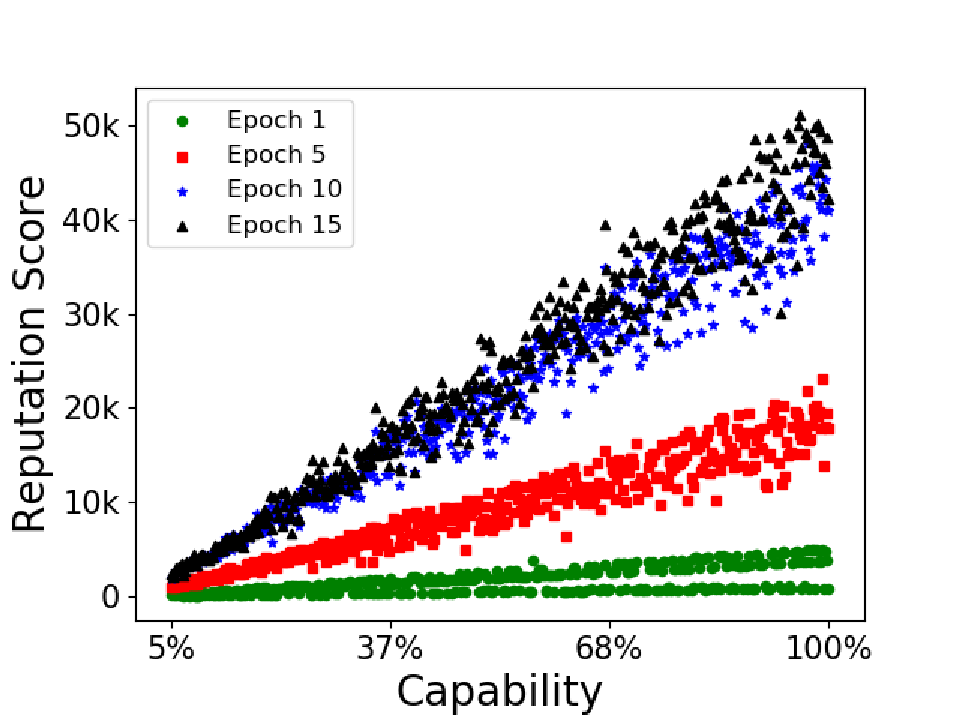}
		\caption{The reputation score of validators with different capabilities from 5\% to 100\%}
		\label{fig:repdistribution} 
	\end{minipage}
	\hspace{0.15ex}
	\begin{minipage}[t]{0.235\textwidth}
		\centering
		\includegraphics[width=1\textwidth]{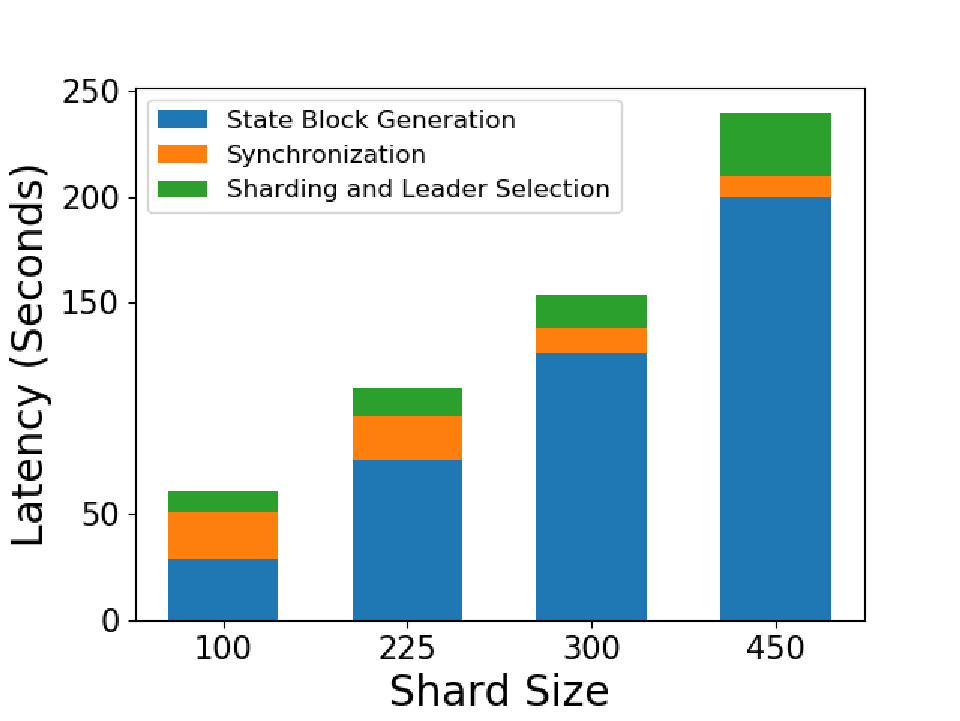}
		\caption{The epoch transition latency}
		\label{fig:transitiontime} 
	\end{minipage}
	\vspace{-4mm}
\end{figure*}    

\subsection{Performance Analysis}
\label{RSthroughput}
Without loss of generality, we analyze the complexity of consensus per $tx$, supposing one $tx$ is processed in the \sys\ of shard size $m$.

Firstly, the time complexity for Raft consensus is $\mathcal{O}(m^2/b)$. At first, the client sends the $tx$ to all the input and output shards. Specifically, the client sends $tx$ to one validator of each shard, and the validator helps broadcast $tx$ within the shard, which requires $\mathcal{O}(m)$ communication cost.
Next, the shard leader generates a TxList of size $b$, and broadcasts both the TxList and $tx$s to all shard members. Such a process brings another communication overhead of $\mathcal{O}(m)$.
Then, the leader collects all TxDec which needs $\mathcal{O}(m/b)$ complexity and broadcasts the TxDecSet which needs $\mathcal{O}(m^2/b)$ time to all validators to reach an intra-shard consensus. In total, it costs  $\mathcal{O}(m^2/b)$ time.

Secondly, the time complexity of a modified RapidChain's consensus is $\mathcal{O}(m^2/b)$. It is reported that the RapidChain's consensus is $\mathcal{O}(m^2/b)$ and the addition \emph{response} round brings a communication overhead of $\mathcal{O}(m)$.
In addition, if $tx$ is a cross-shard transaction, the time will be increased due to the cross-shard communication. The complexity of the increased time is the same as the Omniledger\cite{kokoris2017omniledger} which is $\mathcal{O}(n)$ reported by RapidChain\cite{zamani2018rapidchain}. Therefore, the time complexity of the consensus is $\mathcal{O}(m^2/b+n)$.


It is worth mentioning that the shard leader contributes more bandwidth and computing resources to generate the TxList, TxDecSet, and TB than other validators. Different from the previous work, our scheme explicitly considers capability differences among the validators. Specifically, the reputation scheme eases the workload of the least capable validators to improve the system performance by encouraging the honest and competent validator to contribute more. These validators generally have higher computing resources and bandwidth to process the transactions, thus they will accumulate reputation scores more quickly than others. In our scheme, these validators have more chance of being shard leaders. With more contributions from the more capable shard leaders, the throughput of the system can be significantly improved. 

\subsection{Incentive Mechanism}
\label{RBincentive}
\sys\ provides great incentives, including reputation scores and money - transaction fees to make sure being honest is more beneficial to the validators than being dishonest. In transaction verification, half of the transaction fee is given to the leader and the rest is allocated to the other validators based on their reputation scores earned in the current epoch.
Thus, even a newly joined node could earn reputation scores and get a stable income under our scheme, if it is honest and works hard. 
A malicious node may try to cheat the system by being dishonest occasionally. However, on the one hand, such a node would obtain fewer reputation scores than the honest majority, thus it has barely any chance of being a leader and threatening the system. On the other hand, it also earns much fewer rewards than others during this process, i.e., cheating occasionally is not a good strategy. Thus, it is believed that a rational node would not adopt being dishonest. 
Furthermore, a proper sliding window $w$ is utilized to make sure the selected leader is the honest node who has contributed continuously. In the meantime,  $w$ should not be too large to prevent the monopoly. 
The considerable profit for being the leader would encourage the validators to stay and contribute.

\section{Evaluation}
\label{Evaluation}
In this section, we first describe the setup of our system and then illustrate the detailed evaluation as follows:
\begin{enumerate}[noitemsep]
    \item The performance of \sys\ under different settings.
    \item The scalability of \sys.
    \item The performance under different threat models
    \item The throughput enhancement via utilizing the heterogeneity among validators.
    \item The epoch transition time.
\end{enumerate}
\subsection{Evaluation Setup}

We implement \sys\ in Go language~\cite{golang}. For collective signing, we use the code in Go cryptography libraries~\cite{gocosi}. For the parameters in \sys, we set the sliding window $e$ to be 10. The size of the transaction block is 4MB. No more than a 1/3 of the nodes are malicious nodes. The scaling factor of $S(j)$ in reputation score calculation is set to be 0.1 for a correct decision, 0 for \emph{Unknown}, -0.5 for an incorrect decision where the transaction should be passed but the validator decides to select \emph{Reject}, and -1 for an incorrect decision where the transaction should be rejected but the validator erroneously selects \emph{Accept}. To simulate the transactions, each validator will generate transactions by themselves and send them to each other. 

We run all the experiments on the Amazon Web Service. More specifically, 900 c4.large EC2 instances from two different regions to simulate 1800 nodes. Elastico set all their instances in the US West, Omniledger and RapidChain set all their nodes in one location. Different to them, we set 450 instances in Oregon from the US West and 450 instances in North Virginia from the US East to simulate a realistic, globally distributed deployment. Each instance is equipped with 2 Amazon vCPUs and 3.75GB of memory. The bandwidth is set to be 20Mbps for each node. Every node has the same capabilities in most tests except in the throughput enhancement test. The $\Delta$ is set to be 1.2s according to the TCP ping testing on AWS measured by XFT\cite{liu2016xft}.
\subsection{Performance}
\label{eva:performance}
In this evaluation, we test the performance under different shard sizes of \sys. The metrics of performance comprises two parts: throughput and user-perceived latency. The definition of these two metrics are the same as other works\cite{luu2016secure,kokoris2017omniledger, zamani2018rapidchain}.
The throughput is the number of transactions the system processes in one second. The user-perceived latency is the time that a user sends a $tx$ to the network until the time that $tx$ can be confirmed by any (honest) node in the system\cite{zamani2018rapidchain}. In our case, it is the time when a transaction is proposed until the RB is built so that any user can validate the transaction.
More specifically, the system is tested under the \emph{simple attack} with three different shard sizes: 100, 225, 300 and 450. 
The average throughput and user-perceived latency have been measured and are shown in Fig.~\ref{fig:TPCshardsize} and Fig.~\ref{fig:delayshardisze}.
The throughput for the four shard sizes is 15421 tps, 6853 tps, 4288 tps and 1485 tps respectively. The user-perceived latency is 20.4s, 58.2s, 88.5s, 146s With the increase in sharding size, the throughput will decrease and the latency will increase. 
The poor performance under the shard size of 450 is due to the following reasons. First, \sys\ shifts some workload from the members to the leader.  With the larger shard size, the bandwidth of the leader for package transmission will increase a lot. Second, the CoSi will cost more when the shard size increases.
Anyway, when satisfying the security property, a shard size of 225 has a larger throughput than other previous works. The latency is almost equal to Omniledger. This indicates our design really can support a high-throughput transaction chain.

\subsection{Scalability}
In this section, we test the scalability of \sys. The evaluation measures \sys's throughput based on the shard size of 225 with different numbers of nodes: 450, 900, 1350 and 1800. For each setting, half of the nodes are set in the US West and the remaining half are set in the US East.
The result is presented in Fig.~\ref{fig:TPCscalability}. The throughput is 1834 tps, 3610 tps, 5333 tps and 6853 tps. With more nodes, the throughput  increases linearly. The result indicates the system can be scaled up with more computation power joining the system.

\begin{table*}[]
	\caption{Comparison between \sys\ and the existing works. }
    \begin{tabular}{ccccccclcl}
               & \#Nodes & Resiliency      & \begin{tabular}[c]{@{}l@{}}Complexity\\ (per $tx$)\end{tabular} & Throughput & Latency & \begin{tabular}[c]{@{}c@{}}Shard\\ Size\end{tabular} & \begin{tabular}[c]{@{}l@{}}Balanced\\ Sharding\end{tabular}        & \begin{tabular}[c]{@{}c@{}}Incentive\\ Mechanism\end{tabular} & Time to Fail                                                              \\ \hline
    Elastico   & 1600  & t\textless{}n/4 & $\Omega(m^2/b+n)$         & 40 tx/s    & 800s    & 100      & No                       & No                  & 1 hour                                                                    \\
    OmniLedger & 1800  & t\textless{}n/4 & $\Omega(m^2/b+n)$         & 3500 tx/s  & 63s     & 600      & No                       & No                  & 230 years                                                                 \\
    RapidChain & 1800  & t\textless{}n/3 & $\mathcal{O}(m^2/b+m\log n)$         & 4220 tx/s  & 8.5s    & 200      & \begin{tabular}[c]{@{}l@{}}Simple -\\ active/inactive\end{tabular} & No                  & 1950 years                                                                \\
    RepChain   & 1800  & t\textless{}n/3 & $\mathcal{O}(m^2/b+n)$         & 5628 tx/s  & 58.2s   & 225      & \begin{tabular}[c]{@{}l@{}}Based on \\ reputation\end{tabular}      & Yes                 & \begin{tabular}[c]{@{}l@{}}Depends on attacker's \\ strategy. The worst \\ case is the same as \cite{zamani2018rapidchain}.\end{tabular}
    \end{tabular}
    \label{tab:compare}
    \vspace{-5mm}
\end{table*}
\subsection{Security}
In this section, we test three different attack models: \emph{Simple Attack}, \emph{Camouflage Attack} and \emph{Observe-Act Attack} with 1800 nodes and a shard size of 225. The throughput is measured and shown in Fig.~\ref{fig:security}. The first two attacks can be implemented easily. However, the performance under \emph{Observe-Act Attack} depends on the ability of the attacker. Thus we test our system with a very strong attacker that can observe the score distribution of all the validators. In other words, we cannot distinguish the attackers and honest validators based on their reputation scores.
For the \emph{Simple Attack}, our system runs poorly for the first two epochs, but all the attackers will have a lower reputation after the third epoch so that they can barely degrade the performance of \sys.
For the \emph{Camouflage Attack}, the average throughput is 6104 tps. Once a malicious node becomes leader, the view-change protocol will be launched and its reputation score will be cleared. Thus, the malicious node needs a long time (around 10 epochs) to gain enough reputation score and be selected as a leader again. Based on this, the system can still perform well.
For the \emph{Observe-Act Attack}, the average throughput is 5628 tps which is worse than the previous two attack models. As we mentioned in Sec.~\ref{EpochSecurity}, such an attacker will degrade the \sys\ the same as the random sharding system. In other words, the possibility of a malicious leader is around 1/3 in every epoch. 
From the results and analysis in Sec.~\ref{EpochSecurity}, we can conclude that the \sys\ can enhance the performance and security level when facing a \emph{Simple Attack} and a \emph{Camouflage Attack} and is the same level as the random sharding scheme under \emph{Observe-Act Attack}.

\subsection{Throughput Enhancement}
In this section, we illustrate that the reputation scheme can provide the benefit in throughput enhancement mentioned in Sec.~\ref{RSthroughput}. We set different validators with different capabilities and test whether the reputation scores are relevant to their capabilities. Specifically, we define the capability of the node in Sec.~\ref{eva:performance} as 100\% and limit the speed of handling transactions of the validators so that a low capability validator will give more \emph{Unknon} responses than a high capability validator. In this setting, we set $k=8$ and $n=1800$. Moreover, all the validators are honest in their controlling of the variables and their capabilities follow a uniform distribution from 5\% to 100\%.  
Fig.~\ref{fig:BandwidthTPS} shows the throughput enhancement benefits from our reputation scheme. It can easily be concluded that at the beginning, the throughput is less than 4000 tps due to the low capability leader. However after about 8 epochs, the throughput is averaging at 5243 tps due to a more capable leader being selected based on its cumulative reputation score. We also run 20 epochs of \sys\ without the reputation scheme, \textit{i.e.}, the validators are randomly sharded into 8 groups and the leader is randomly elected. The average throughput is 3512 tps which is similar to the first epoch but much lower than the following epochs of \sys\ with the reputation scheme.
Fig.~\ref{fig:repdistribution} shows the relationship between their capabilities and the reputation score in the 1st, 5th, 10th and 15th epochs. We can see the validators barely distinguish between each other at the first epoch. However with more epochs, their abilities can be more easily distinguished according to their reputation scores. Also, the relationship between reputation scores and their capabilities is linear, which indicates the reputation score can correctly reflect the ability of the validators. 
From the results above, \sys\ can distinguish the validators with different capabilities and select a better leader via the sharding and leader selection scheme. Thus, our design can enhance the throughput of the sharding-based blockchain system and incent the highly capable nodes.

\subsection{Epoch Transition Latency}
In this evaluation, we test the average epoch transition latency of \sys\ under different shard sizes of 100, 225, 300 and 450. The latency includes the time for generating state block via collective signing, PoW, synchronization, sharding and leader selection. The results are shown in Fig.~\ref{fig:transitiontime} which is 61.2s, 111.8s, 155.3s and 242.0s. Among the three stages, the state block generation costs the most time. With a larger shard size, the transition latency will increase due to the overhead on collective signing. 

\subsection{Compare with other works}
In this section, we compare \sys\ with other sharding blockchain works. Monoxide\cite{wang2019monoxide} are built on PoW which is different with other BFT based works. The work of \cite{dang2019towards} is reliant on SGX which needs dedicated hardware.
Thus we list the main metrics of remaining works\cite{luu2016secure,kokoris2017omniledger, zamani2018rapidchain} as Table.~\ref{tab:compare} shows.

We argue that compared with the above works, our work integrates reputation with a sharding-based blockchain, and explicitly leverages the benefits brought about by reputation on incentive scheme, throughput and security level to fully realize the potential of the sharding technique.

In terms of throughput, existing works ignore capability differences between validators so that less-competent ones may hinder the overall performance. Our system leverages reputation to characterize such heterogeneity and elects the competent nodes as shard leaders to help improve throughput of the raft-based consensus. It is worth mentioning that the time complexity of Rapidchain is the smallest among all the works. However, the cost is to generate at least three more transactions for one cross-shard transaction. Considering the majority of the transactions are cross-shard transactions, such cost means only one fourth of their throughput is used for the transaction sent by the user.

In terms of incentive mechanism, some works~\cite{ kokoris2017omniledger, zamani2018rapidchain, dang2019towards} do not consider the incentive mechanism. Other works~\cite{luu2016secure,wang2019monoxide} only give rewards to block miners, which causes large income variance for miners with modest computational power~\cite{luu2017smartpool}. RSCoin~\cite{danezis2015centrally} resorts to a central bank, which is undesirable in most blockchains. 
Our system allocates rewards based on validators' reputations, which recognizes every devoted contribution and incents validators to do their best. Compared with RepuCoin~\cite{yu2019repucoin}, our reputation scheme is integrated with decays property to alleviate the monopoly problem. 

In terms of security, almost all these systems adopt random sharding and leader selection except RapidChain~\cite{zamani2018rapidchain}. RapidChain proposes a coarse committee reconfiguration built on Cuckoo rule, which simply regards half of the validators as active nodes and those remaining as inactive nodes. Our scheme further utilizes reputations to establish a more balanced (\textit{i.e.}, each shard has similar proportions of active, inactive, honest and malicious validators) sharding scheme, so that it is difficult for an attacker to take control of one shard. Sec.~\ref{Analysis} shows that the time to fail of \sys\ is the same as RapidChain under the same $n$ and $m$ in the worst case - \emph{Observe-Act Attack}.







\section{Discussion}
\label{Discussion}
In this section, we discuss the future works.

\subsection{Information Dispersal Algorithms}
In \sys, all the validators send messages to each other via TCP. Such a setting results in a relatively high latency that reduces the system throughput. However, it can be improved by leveraging more efficient information dispersal algorithms, such as IDA-Gossi, an unreliable broadcast protocol with an erasure code scheme proposed by RapidChain\cite{zamani2018rapidchain}.

\subsection{Slowly Adaptive Attacker}
Slowly adaptive attacker wherein the attacker determines the
victim between epochs but cannot change this set within the epoch is a common threat model in other sharding blockchain systems~\cite{kokoris2017omniledger,luu2016secure,zamani2018rapidchain}. However, such an attacker could easily defeat a reputation based blockchain system such as \sys\ as well as other state-of-the-art works\cite{huang2019b,yu2019repucoin,bahri2019trust}. Such attacker seems too strong since they consider all the victims to have the same security level. In this paper, we consider a more realistic model where different validators may have different security levels. Thus, we consider a more realistic threat model in Sec~\ref{Overview}. In future work, we would like to solve this problem by adopting some privacy-preserving technologies such as zero-knowledge proof.

\subsection{Dynamic Parameters}
In the worst case of the current \sys, an advanced \emph{observe-act attack} degrades the system security to the same as the random-based solutions. However, we can improve it via dynamically adjusting some system parameters. Specifically, \sys\ could adjust the reputation score formula, the sliding window and the sharding scheme dynamically based on some randomness. Therefore, the distribution of reputation scores would change in every epoch with some randomness, which effectively defends the observe-act attacks.


\section{Conclusion}
\label{Conclusion}
This paper proposes \sys, a reputation-based secure, fast, and high incentive blockchain system via sharding. \sys\ leverages reputation scores which describe the heterogeneity among validators and provide the incentive mechanism. \sys\ is the first sharding blockchain with double-chain architecture. For transaction chain, a Raft-based synchronous consensus which achieves high throughput has been presented. For reputation chain, a collective signing has been utilized to achieve a consensus on the reputation score and support the high throughput transaction chain with a moderate generation speed. To boost the system throughput and enhance its security, a reputation based sharding and leader selection scheme has been proposed. To analyze the security of \sys, we propose a recursive formula to calculate the epoch security within only $\mathcal{O}(km^2)$ time. The evaluation shows \sys\ have a good performance under different threat models. It can also enhance both the throughout and security level and provide an incentive mechanism to the existing sharding-based blockchain system.

\bibliographystyle{IEEEtran}
\bibliography{IEEEabrv,reference.bib}
\end{document}